\documentstyle[aps,preprint,citesort,epsf,fancyhdr]{revtex}

\newcommand{\secref}[1]{Section~\protect\ref{#1}}
\newcommand{\eqref}[1]{Eq.~(\protect\ref{#1})}
\newcommand{\figref}[1]{Fig.~\protect\ref{#1}}

\begin{document}

\lhead{}   
\chead{}
\rhead{}
\lfoot{}
\cfoot{Stationarity and Redundancy of\ldots \hfill
Zoldi \hspace{.3in} \today \hfill Page~\thepage}
\rfoot{}
\renewcommand{\headrulewidth}{0pt}
\renewcommand{\footrulewidth}{0.3pt}

\draft

\pagestyle{plain}

\begin{center}
\Large\bf
Stationarity and Redundancy of Multichannel EEG Data
Recorded During Generalized Tonic-Clonic Seizures
\end{center}

\vspace{.2in}

\begin{center}
\large
Scott M.~Zoldi, PhD\\
T-CNLS, MS B258\\
Los Alamos National Laboratory\\
Los Alamos, NM 87545; zoldi@cnls.lanl.gov\\
\ \\
Andrew Krystal${}^\dag$, M.D.\\
Department of Psychiatry\\
Duke University Medical Center, Box 3309\\
Durham, N.~C. 27708; krystal@phy.duke.edu\\
\ \\
Henry~S.\ Greenside${}^\dag$, PhD\\
Department of Physics, Duke University\\
Durham, NC 27708-0305; hsg@phy.duke.edu
\end{center}

\vspace{.2in}

\begin{center}
\large
May 4, 1999
\end{center}

\newpage
\pagestyle{fancy}

\begin{abstract}

A prerequisite for applying some signal analysis methods to
electroencephalographic (EEG) data is that the data be
statistically stationary. We have investigated the
stationarity of 21-electrode multivariate EEG data recorded
from ten patients during generalized tonic-clonic (GTC)
seizures elicited by electroconvulsive therapy (ECT).
Stationarity was examined by calculating probability density
functions (pdfs) and power spectra over small equal-length
non-overlapping time windows and then by studying visually
and quantitatively the evolution of these quantities over
the duration of the seizures. Our analysis shows that most
of the seizures had time intervals of at least a few seconds
that were statistically stationary by several criteria and
simultaneously for different electrodes, and that some leads
were delayed in manifesting the statistical changes
associated with seizure onset evident in other leads.  The
stationarity across electrodes was further examined by
studying redundancy of the EEG leads and how that redundancy
evolved over the course of the GTC seizures.  Using several
different measures,
we found a substantial redundancy which suggests that fewer
than 21~electrodes will likely suffice for extracting
dynamical and clinical insights.  The redundancy analysis
also demonstrates for the first time posterior-to-anterior
time delays in the mid-ictal region of GTC~seizures, which
suggests the existence of propagating waves.  The
implications of these results are discussed for
understanding GTC seizures and ECT treatment.
\ \\
\ \\
Key words: ECT, generalized seizures, stationarity, redundancy.

\end{abstract}

\newpage
\section{Introduction}
\label{intro}
The analysis of multilead electroencephalographic (EEG) time
series is an important but difficult goal. Such an analysis
is important because the EEG provides a relatively low-cost,
noninvasive way to monitor brain behavior that can yield
valuable insights about brain function, pathology, and
treatment. The goal is difficult because of the great
complexity of brain dynamics which remains poorly
understood.

For EEG data analysis, the complexity of brain dynamics
manifests itself in at least three different ways. First,
EEG time series are nonstationary which makes difficult the
application and interpretation of many methods of signal
analysis including those based on
statistics~\cite{Johnson92} and nonlinear
dynamics~\cite{Abarbanel93}. It is not understood yet how to
quantify the magnitude and time-dependence of
nonstationarities, how to compare the nonstationarity of one
EEG recording with another, or how to identify regions of
approximately stationary behavior.  Second, multielectrode
EEG data are spatiotemporal in character since the time
series are recorded simultaneously from electrodes at
different locations on the scalp.  Here the relation between
time series at different electrodes is not well understood,
e.g., to what extent is there redundancy of the time series
from different electrodes or how does one interpret the
correlations that may exist.  A third difficulty is the
statistical variability of EEG data, which can be
substantial even for EEG recorded from the same patient
under presumably similar conditions. This variability
complicates the extraction of features that can be used to
compare one EEG with another or to classify EEG for clinical
purposes.

In this paper, we apply and compare several statistical and
visualization methods to quantify the stationarity and
redundancy of 21-electrode EEG time series measured during
\underline{generalized tonic-clonic (GTC)} seizures
associated with \underline{electroconvulsive therapy (ECT)}
treatments. Such seizures are characterized by polyspike and
slow-wave activity~\cite{Weiner91} and are often observed to
be nonstationary. Although
nonstationarity~\cite{Kawabata73,Cohen77,Barlow85,Ferber87,MayerKress94}
and
redundancy~\cite{Pfurtshceller71,Pijn90,Silva89,Pijn90,Mars82}
of EEG data have been studied previously, this earlier work
has not addressed GTC seizures.  ECT provides a unique
opportunity for the study of GTC seizures because the
seizures are induced under controlled conditions. The
patient is anesthetized and given a neuromuscular relaxant
(succinylcholine) which eliminates movement artifacts that
would otherwise obscure the EEG data. The seizure is induced
using a reproducible procedure in which a current of known
amplitude and wave form is administered between two
electrodes attached to particular locations on the
scalp~\cite{Krystal96}. We partly address the issue of
statistical variability of the EEG time series by recording
and comparing ECT-induced GTC seizures from ten patients.

Our analysis of EEG data associated with GTC seizures has
two goals. One is to further basic science by understanding
more clearly when EEG signals may be considered stationary
so that particular signal analysis methods might be applied
more successfully. Our results for GTC ECT-related seizures
show that several commonly used criteria for stationarity
such as a \underline{power spectral density (psd)} and
\underline{probability distribution functions (pdfs)}
applied to successive windows of data often do identify the
same stationary parts of the data but there are
qualifications as we discuss in
\secref{results-stationarity} below.  In analyzing EEG
redundancy, our results further suggest that the majority of
electrodes are often stationary or nonstationary
together. There is a high cross-correlation between
electrodes (see \eqref{rxy-eq} below), which has been
observed in previous EEG~research but not quantified in the
context of GTC seizures~\cite{Silva89}.

A second goal of this paper is to understand more
specifically the electrophysiology of GTC ETC seizures. The
concept of ``seizure generalization'' is used frequently to
describe spatial aspects of ECT seizures and has been
considered to be a central factor in determining the
therapeutic effectiveness and side-effects associated with
ECT treatments~\cite{Swartz86,Abrams91}.  For example, it
has been suggested that marginal seizures are less
generalized spatially~\cite{WeinerKrystal93}.  Several
authors have suggested that the differences in efficacy and
side-effects of \underline{unilateral~(UL)} and
\underline{bilateral~(BL)} ECT (the two most commonly
employed electrode placements) are due to differences in
generalization~\cite{Swartz86,Abrams91,Swartz95}.  Although
ECT seizures have been considered to be generalized
tonic-clonic seizures~\cite{Niedermeyer93}, we observe a
variation in generalization through spatial inhomogeneities
of the EEG which suggest that generalization is a graded
phenomenon (not all or nothing).



Our data analysis leads to new conclusions. We quantify
earlier claims~\cite{Weiner91,Staton81,Weiner83} that the
mean frequency of an ECT seizure decreases steadily during
the seizure. (This has recently also been reported with
GTC~seizure EEG data using a different set of techniques
based on time-varying autoregressive
modeling~\cite{West99}.) We also identify statistical
differences in ECT seizures generated by unilateral and
bilateral electrode stimulation. Our statistical analysis
based on time delays also provide the first evidence for
wave propagation during ECT seizures. We identify evidence
that seizure activity is expressed regionally in the brain
and that some regions are delayed in manifesting the
statistical changes characteristic of seizure activity in
other leads.  This variation, particularly in the temporal
and pre-frontal regions, may have implications for
understanding the variable cognitive side-effects and
anti-depressant efficacy associated with the induced
seizures.

The rest of the paper consists of the following sections. In
\secref{prior-work}, a survey of prior work is given. In
Sections~\ref{clinical-methodology} and~\ref{quant-methods},
details are discussed concerning how the ECT EEG data were
clinically recorded and analyzed. In \secref{results}, we
discuss our results on stationarity and redundancy. Finally,
in \secref{conclusions}, we summarize our conclusions and
discuss some questions for further study.

\section{Previous Work}
\label{prior-work}
In this Section, we review prior work with an emphasis on
stationarity and redundancy. In the following Section, we
discuss details of the methods that we use to measure and to
analyze the ten-patient ECT~EEG data set recorded at Duke
University's Quantitative~EEG Laboratory.

\subsection*{Stationarity}

Researchers have studied EEG stationarity with several
methods and some of this earlier work motivated our own
analysis. One method involves partitioning time series into
equal-size non-overlapping segments (typically a few seconds
long), calculating power spectra for each segment, and then
studying how the power spectra evolve from one segment to
the next~\cite{Kawabata73,Barlow85,Ferber87}.  In another
EEG study of patients in the eyes-closed waking state,
researchers have compared the means of time series over
successive segments and found that segments shorter than
12~seconds could often be considered stationary by this
criterion~\cite{Cohen77}.  Stationarity has also been
studied in EEG data during sleep by testing whether there
were trends over time in the signal
variance~\cite{Sugimoto78}.  Still another approach has been
to compute and to compare probability distributions (pdfs)
of the EEG signal over time~\cite{Sugimoto78,Barlow85}.

\subsection*{Redundancy}

Many of the studies of stationarity mentioned above concern
EEG data from only one electrode and direct observation of
EEG time traces indicates that not all EEG channels are
stationary or nonstationary together.  Although correlations
and redundancy have been studied for focal
epilepsies~\cite{LeVanQuyen99}, researchers have not yet
studied these for generalized
seizures~\cite{Mars82,Pijn90,Silva89}. Instead, several
authors who have studied ECT seizures have noted spatial
inhomogeneities with a tendency for the EEG~amplitude to be
greatest in the central part of the
scalp~\cite{Enderle86,Weiner91}. An amplitude asymmetry in
which electrodes on the right side have larger amplitude
than those on the left has been described for right
unilateral ECT (in which the stimulating electrodes are
placed at the vertex and right temple) but this has not been
observed for bilateral ECT (in which the electrodes are
placed bi-temporally)
\cite{Small70,Staton81,Weiner91,Krystal92,WeinerKrystal93,Krystal93}.
Researchers have also observed delays in the onset of
seizure activity of about a second when comparing time
series from two different leads~\cite{Staton81,Brumback82}.
 
Some preliminary studies have analyzed the inter-hemispheric
redundancy of ictal EEG data but only in two-channel data,
for electrodes Fp1 and~Fp2 referenced to the ipsilateral
mastoid.  In the 2-5~Hz frequency band, greater coherence
(defined below in \eqref{rxy-eq}) in the first 6~seconds
after the stimulus and a lower coherence in the 6~seconds
immediately after the end of the seizure have been
associated with a greater likelihood of ECT therapeutic
benefit~\cite{Krystal95,Krystal96}.

Several previous studies have also employed techniques for
quantifying the spatial redundancy of EEG data.  Such
techniques include linear
correlation~\cite{Pfurtshceller71,Pijn90}, nonlinear
correlation~\cite{Silva89,Pijn90}, the average amount of
mutual information~\cite{Mars85,Silva89,Pijn90},
coherence~\cite{Gotman81,Leuchter92}, and estimates of
time-delay based on these
measures~\cite{Gotman81,Mars85,Silva89,Pijn90}. Motivated by
these reports, we have studied interlead linear correlation,
average amount of mutual information, and interlead time
delays based on these two measures.

\section{Clinical Methodology}
\label{clinical-methodology}
In this section, we discuss the experimental details of
obtaining 21-electrode ECT EEG data.  As a first step, ten
subjects were studied who had been clinically referred for
ECT. These subjects consisted of 6~women and 4~men with ages
ranging from 45-73 years, representing a typical clinical
ECT~population. Prior to each treatment, the barbiturate
methohexital and the muscle relaxant succinylcholine were
administered at a dosage of 1~mg/kg according to standard
ECT practice~\cite{APA90}.  All subjects were free of any
antidepressant, anticonvulsant, antipsychotic, or mood
stabilizing medications for at least 5~days prior to the ECT
course.  A single seizure was recorded for each subject.

Experiments have shown that the dynamics and clinical
benefits of an ECT seizure depend on the placement of the
ECT electrodes on the scalp (e.g., UL or~BL) and on the
stimulus intensity as measured in units of the threshold
intensity needed to induce a
seizure~\cite{Sackeim91,Sackeim93,Nobler93,Krystal95,Krystal96,Krystal97,Krystal99}.
BL ECT may be more effective, and higher-above-threshold
stimuli appear to be more efficacious, particularly for
UL~ECT.  Of the ten subjects studied in this paper, eight
received pulse right unilateral ECT~\cite{dElia70} and two
received bilateral ECT. We note that the stimulus electrode
placement was not controlled in this study but was
determined clinically and happened to include more
UL~subjects. Nonetheless, this study allows a preliminary
comparison of dynamical EEG differences in these two forms
of treatment. Electrical stimulus dosing was administered
via a standard clinical technique in which the seizure
threshold at treatment~1 was determined and then a stimulus
at subsequent treatments was delivered that was a multiple
(in terms of charge) of the determined
threshold~\cite{Weiner94}.

During the ECT treatment, twenty-one channels of EEG data
(nineteen EEG leads plus two eye leads) were recorded using
the International 10/20 System with locations indicated as
in \figref{fig:electrode-location}.  Because of the
placement of ECT stimulus electrodes, position F5 was used
(midway between positions~F7 and~F3) as was~F6 (midway
between~F8 and~F4) but for convenience we refer to~F5 and~F6
throughout as leads~F7 and~F8. All leads were referenced to
linked ears and recorded using Ag/AgCl electrodes.  The data
were amplified and filtered using a Nihon-Khoden 4221 device
(Nihon-Khoden Corp.) with a low-frequency cutoff of 1.6~Hz
and a high-frequency cutoff of 70~Hz.  The data were
digitized at 256~Hz with 12-bit accuracy in the form of
integers between the values -2048 and 2048.  Although the
use of succinylcholine greatly diminishes artifacts in
the~EEG that would otherwise be present, some artifacts may
occasionally still occur.  As a result, the second author,
A.D.K., carefully screened all EEG data for artifacts.
Brief segments of EEG data from 2~subjects were excluded
from analysis on this basis.

\section{Methods for Quantifying Stationarity and Redundancy}
\label{quant-methods}

   In this section, we discuss the methods that we use to
analyze the ECT EEG data for stationarity and redundancy. We
first discuss stationarity, for which various quantities are
calculated over successive equal-sized non-overlapping time
windows and over each electrode of the multivariate
recording.  We next discuss measures for quantifying
redundancy of the multivariate EEG data and how this
redundancy may evolve in time. We conclude this section by
discussing how one can study whether stationarity and
redundancy are affected by time-delays between a given pair
of electrodes.

We note that our emphasis in using these methods is somewhat
different than that of a statistician or of a nonlinear
dynamicist. Since GTC~seizures terminate spontaneously 1-2
minutes after induction, the corresponding EEG time series
are obviously nonstationary.  There are then three
interesting questions.  One is simply how to characterize
the nonstationary dynamics so as to provide insights into
the properties of a GTC seizure and this is the main
emphasis of our work. A second question is whether there are
significant windows of approximately stationary behavior
which one could then treat by statistical and nonlinear
techniques that assume stationarity. A third question, which
we do not address here, is whether some hidden component of
the EEG signals are statistically stationary, e.g., whether
a particular filtering of the data would yield stationary
behavior.  Given the complexities of brain physiology and of
the time series themselves (there is no known underlying
mathematical model based on first principles of neuronal
physiology), our analysis should be regarded as exploratory
rather than intended to falsify specific hypotheses of the
statistical structure of ECT EEG data.

   \subsection{Measures of Stationarity}
   \label{stationarity-measures}
   A statistical process is defined to be {\em stationary} if
its statistical properties are time-translation invariant,
i.e., shifting the origin of time (making the substitution
$t \to t + t_0$ where~$t_0$ is some constant) has no effect
on the statistics of the process. (A weaker definition of
stationarity is a process for which only the mean and
variance of the process are established to be
time-translation invariant.) A statistical process is {\em
nonstationary} if any of its statistical properties depend
on time~\cite{Priestley88}.

Although not often stated explicitly, it should be
appreciated that the definitions of stationarity and
nonstationarity involve mathematical idealizations and so
are impossible to establish rigorously with a finite amount
of empirical data. Technically, an infinite amount of data
is required to define a statistical property such as a
probability distribution or joint probability distribution.
Similarly, one needs infinitely many realizations to study
statistical properties over some interval of time and then
to examine whether those properties change with the choice
of time interval.


Besides the challenge of testing the hypothesis of
nonstationarity with a finite amount of data, there is a
related conceptual difficulty that there is no natural
measure of ``degree of nonstationarity'' or ``magnitude of
nonstationarity''. This is a simple consequence of the fact
that nonstationarity is defined as the negation of
stationarity, so that an arbitrarily weak time dependence of
any statistical property is enough to make a time series
nonstationary. One then has to look carefully at specific
data and hope that certain features will suggest themselves
as significant for causing nonstationary statistics.

Our approach for testing nonstationary structure was to
divide all time series into successive non-overlapping
equal-sized time windows (also called epochs), calculate
some statistical properties of the data in each window, then
study these statistical properties as a function of time
(from one window to the next).  Each window was chosen to be
1- or 2-seconds in length, with the length chosen
qualitatively after examining visualizations of several
statistical quantities as a function of time (see
\figref{fig:global-variance}(a) as an example). These time
intervals were sufficiently long to contain enough points
(256 and 512~points respectively for~1- and 2-second
windows) for reasonable estimates of statistical quantities
yet were empirically short enough that time variations of
the statistics could be examined over the typical
0.5-2~minute duration of an~ECT seizure.  Our results were
weakly dependent on the window width over a range of
1-4~secs. In future studies, it would be useful to explore
some recently proposed stationarity tests that avoid the use
of windows~\cite{West99}.

Once a window length was determined, we used three
statistical quantities to monitor possible nonstationary
behavior: a window variance~$\sigma^2$, the mean
frequency~$\langle \omega \rangle$ of the power
spectrum~$P(\omega)$ calculated over the time series in the
window, and a $\chi^2$ statistic that measured the deviation
of the probability distribution function (pdf) $\rho(x)$ in
a given window (where~$x$ denotes the amplitude) from a
cumulative pdf based on the time series of previous
approximately stationary regions.  In our analysis, windows
were the same size and synchronized across all EEG
electrodes and so each of the 19~time series of a particular
EEG recording produced three new shorter time series
representing the window-dependence of the above three
statistical quantities. To visualize statistical trends
across all 19~electrodes, these shorter time series were
next plotted as a matrix of color pixels~$M_{ij}$ by
assigning a color palette to the range of the shorter time
series. Each row of the matrix indicates the time dependence
(from left to right) of a statistic associated with a
particular electrode (see for example
\figref{fig:global-variance}), while each column represents
the values of a statistic for all electrodes in a given
window.

The three statistical quantities were calculated as
follows. The variance~$\sigma^2$ over a given window was
calculated via the usual statistical formula
\begin{equation}
  \sigma^2 = {1 \over N - 1} \sum_{i=1}^N
       \bigl( x_i - \langle x \rangle \bigr)^2
  , \label{variance-defn}
\end{equation}
where~$N$ is the number of data points in a given window,
$x_i$ are the values of the EEG time series, and $\langle x
\rangle = (1/N)\sum_{i=1}^N x_i$ denotes the average value
of the time series over the window.

The mean frequency $\langle \omega \rangle$ over a given
window was obtained from a frequency-weighted average of the
power spectrum~$P(\omega)$ over that window:
\begin{equation}
  \langle \omega \rangle =
    { \sum_{j=0}^{N/2} \omega_j P(\omega_j) \over
      \sum_{j=0}^{N/2} P(\omega_j) }
  . \label{mean-freq-defn}
\end{equation}
(The sums go from~0 to~$N/2$ because
the power spectrum has only $1+N/2$ separate magnitudes of
Fourier coefficients for a time series of length~$N$.) The
power spectrum~$P(\omega)$ over a given window was estimated
with a Fast Fourier Transform~\cite{Press92}.  Power spectra
for overlapping intervals of length 2~seconds (each
overlapping by 1~second) were averaged to reduce the
variance of the spectrum. Each time series was also
multiplied by a Parzen window before being Fourier analyzed
to reduce artifacts arising from the nonperiodicity of the
time series over the window~\cite{Press92}. The windowing of
the data allowed a frequency resolution in $P(\omega)$ of
$\triangle\omega = 0.5 \, \rm Hz$. We note that the variance
$\sigma^2$ calculated for a window is proportional to the
integral $\int P(\omega) \, d\omega$ of the power spectrum
over the window and so is not entirely independent of the
power spectrum.

The pdf $\rho(x)$ of a time series~$x_i$ in one-second-long
windows was calculated by binning the data (256 points
of~$x_i$)
into 40~bins that spanned the range $[x_{\rm min},x_{\rm
max}]$ of the minimum~$x_{\rm min}$ and maximum~$x_{\rm
max}$ of the time series. Several different numbers of bins
varying from 20 to~200 were studied before establishing
that~40 was adequate in capturing most features of the pdf
without too much statistical noise.

Since it is difficult to plot and to understand the time
dependence of functions like pdfs for multivariate data and
for many different electrodes, the nonstationarity of the
pdfs was analyzed by plotting instead whether each pdf
passed a $\chi^2$ test~\cite{Press92} at the 95\% level
which measured the difference between the pdf in a given
current window and a cumulative pdf over previous contiguous
stationary windows. A cumulative pdf has the advantage of
increasing the statistical accuracy when comparing a new pdf
with a previous standard.  (Recently Witt et
al~\cite{Witt98} also suggested using a $\chi^2$~test for
pdfs to quantify nonstationarity in a time series, but they
did not use a cumulative pdf as we do here.)

The $\chi^2$ value for a particular window was calculated as
follows.  If~$M$ denotes the number of bins (here $M=40$),
and $n_i$ and $N_i$ denote respectively the number of points
in the $i$th bin of the current and cumulative pdfs, then we
calculated the number~\cite{Press92}
\begin{equation}
  \chi^2 = \sum_{i=1}^M { (n_i - N_i)^2 \over (n_i + N_i) }
  . \label{chi-square-eqn}
\end{equation}
Nonstationarity was then visualized by assigning values of~0
and~1 to windows that respectively failed and passed
the~$\chi^2$ test.
A visualization of such $\chi^2$ values as a function of
window index~$i$ is given in \figref{figure:stationarity}
and discussed further below in
\secref{results-stationarity}.

   \subsection{Redundancy and Time Delays}
   \label{redundancy-measures}
   Redundancy of the 19~electrode time series and of statistics
calculated for the 19~electrode time series were quantified
using linear correlation coefficients and mutual
information, with large values of these quantities
corresponding to substantial redundancy. The linear
correlation coefficients~$r_{xy}$ between two time
series~$x_i$ and~$y_i$ were estimated from the sample
correlation coefficient defined as
follows~\cite{Johnson92,Press92}:
\begin{equation}
  r_{xy} = {
   \sum_{i=1}^N \left( x_i - \langle x \rangle \right)
                \left( y_i - \langle y \rangle \right)
   \over
   \sqrt{ \sum_{i=1}^N \left( x_i - \langle x \rangle \right)^2 }
   \sqrt{ \sum_{i=1}^N \left( y_i - \langle y \rangle \right)^2 }
  }
  . \label{rxy-eq}
\end{equation}
In this paper we studied the 18~correlation
coefficients~$r_{x,CZ}$ of all electrodes
with the centrally located electrode~CZ.  The choice of~CZ
was motivated by earlier work~\cite{Weiner91} which showed
that ECT EEG~signals are often largest in amplitude in an
apparently highly correlated region centered around
lead~CZ. The use of this lead in analyses of redundancy of
other leads thus helped to test when leads were highly
related to the dominant activity.  The 18~coefficients
$r_{xy}$ giving the time-evolution of redundancy with
electrode~CZ were computed over a segment of the EEG where
at least 15~electrodes were simultaneously
stationary.

Since it is known from studies in nonlinear dynamics that
linear coefficients such as \eqref{rxy-eq} sometimes miss
nonlinear correlations~\cite{Li90}, we supplemented our
analysis of cross-correlation by studying the mutual
information of pairs of time
series~\cite{Mars82,Fraser89,Li90}. If the quantities
$\rho_x(x)$ and $\rho_y(y)$ denote the pdfs for two time
series $x_i$ and $y_i$ on a given window and if
$\rho_{xy}(x,y)$ denotes the joint pdf (estimated
numerically by sorting pairs of points $(x_i,y_i)$
simultaneously into 40~bins spanning the $x$-range and
40~bins spanning the $y$~range), then the mutual information
$I(x,y)$ is defined to be \cite{Fraser89}:
\begin{equation}
  I(x,y) = \sum_{i=1}^N \sum_{j=1}^N 
    \rho_{xy}(x_i,y_j) \log\left(
      \rho_{xy}(x_i,y_j) \over \rho_x(x_i) \rho_y (y_j)
    \right)
  . \label{mutual-info-eq}
\end{equation}
For statistically independent time series, the joint
distribution $\rho_{xy}(x,y) = \rho_x(x) \rho_y(y)$ factors
into a product of the separate pdfs and $I(x,y)$ becomes
zero.  Empirically we found that 2-second windows contained
enough data to generate good approximate joint probability
distributions; if larger segments of data were studied the
time evolution of redundancy could not be studied.  Mutual
information coefficients between electrode~CZ and the other
18~leads were calculated every 2~seconds to obtain their
time dependence over the seizure.

%

All interlead time-averaged linear correlation and mutual
information coefficients were calculated over ``global''
stationary regions determined by the stationarity tests
indicated above.  These global stationary regions were
identified visually as the largest continuous part of the
time series that was stationary for the majority of
electrodes.  The average interlead coefficients between all
pairs of electrodes were represented in a $19 \times 19$
square matrix.  These matrices allowed us to determine the
average redundancy between one lead and all other leads in
the seizure using both linear correlations and mutual
information (see Figs.~\ref{figure:ave-mutual}
and~\ref{figure:ave-correlation}).

The correlation function \eqref{rxy-eq} and the mutual
information \eqref{mutual-info-eq} were also used to measure
time-delays between two time series $x_i$ and~$y_j$
associated with two different electrodes.  To compute the
time-delay, one time series $y_i$ was fixed and and the
second series~$x_i$ waveform was then shifted in time from
-40~ms to~40~ms (corresponding to integer shifts $x_{i+k}$
of $k=-10$ and $k=+10$) to find the time-delay that resulted
in the maximum value of the mutual information.  We then
determined the shift that gave the largest redundancy
according to \eqref{rxy-eq} or \eqref{mutual-info-eq}.  To
reduce the large number of inter-lead comparisons,
time-delays were again calculated only for leads paired with
the central electrode~CZ. We found 2-second or larger
stationary segments of uniform time-delay in one half of the
seizures. In~\figref{figure:time-delay}, we display the
time-average of the time-delays over a global stationary
region over the surface of the head.

\section{Results and Discussion}
\label{results}

  \subsection{Stationarity}
  \label{results-stationarity}
  \subsubsection{Signal Variance Over Time}

By using plots similar to \figref{fig:global-variance} to
examine all 10~seizures, we found a substantial variation
between seizures in the pattern and degree of stationarity,
as measured by the signal variance.
Figs.~\ref{fig:global-variance}a
and~\ref{fig:global-variance}b are representative of the
diversity of variance evolution that was present across
these seizures. \figref{fig:global-variance}a illustrates
that the variance remains relatively low for the first 18
seconds of the seizure in all leads and then increases (as
indicated by the change from blue and green to yellow and
red) for the fourteen seconds thereafter, but only in the
fronto-central region of the head (leads FP1, F3, FZ, CZ,
FP2, and~F4).  This behavior is typical of the increase in
amplitude that has previously been described in the
transition from the early tonic phase of the seizure to the
later larger amplitude mid-ictal poly-spike and wave EEG
pattern characteristic of the clonic phase of generalized
tonic-clonic seizures, which is largest in amplitude in the
fronto-central region~\cite{Weiner91,Weiner93,Krystal96}.
\figref{fig:global-variance}(a) also manifests two periods
of apparent stationarity of signal amplitude in that for
times 2-16 (a 14-second segment) and times~20-32 (a later
12-second segment) there is minimal change in the color of
the figure in any channel.

In contrast, \figref{fig:global-variance}b is a seizure
whose variance is three times smaller and that has briefer
segments of amplitude stationarity.  There are a number of
approximately 5-second segments that appear to maintain
consistent variance across the head but not longer
stationary segments.  The amplitude is once again greatest
fronto-centrally but only intermittently.

To better illustrate the range of variance stationarity
present, we note that for~8 of~10 seizures, stationary
segments of 8~seconds or longer could be identified where
there was little change in the signal variance in any lead.
The other two seizures had highly variable variance as shown
in \figref{fig:global-variance}b.  The largest stationary
segment present in a single-lead in any of the seizures was
80~seconds.  The largest segment that was stationary across
all of the leads was 30~seconds in length.

There was some consistency across the seizures in the
spatial and temporal patterns of variance.  The
fronto-central leads tended to be larger in amplitude than
the temporal and occipital leads in nine of ten of the
seizures including both unilateral and bilateral seizures.
For most leads the variance increased from the start of the
seizure to the mid-ictal portion, however this increase was
relatively delayed in the onset of the temporal and
occipital leads.  For 6~of the 10~seizures, the time at
which the variance increased was delayed in at least one
lead for both unilateral and bilaterally induced seizures.
In some of the seizures a few leads never manifested an
increase in variance.  This can be seen in
\figref{fig:global-variance}a where T4, T6, T3, T5, P3, O1,
O2, and~F8 do not appear to increase in amplitude and in
\figref{fig:global-variance}b where there does not appear to
be an increase for leads~T3 and~T5.

We conclude that the variance indicates a range of amplitude
stationarity for generalized tonic-clonic seizures.  For
most seizures (eight of ten) a region of stationarity of at
least 8~seconds can be expected for all leads, however there
are some seizures where only much briefer periods of
amplitude stationarity can be found.  These analyses also
indicate that the signal amplitude tends to be greatest, and
that there is an earlier onset in increased variance, in the
fronto-central as compared with temporal and occipital leads
and occasionally frontopolar leads.

\subsubsection{$\chi^2$ Stationarity Test}

\figref{figure:stationarity}(a) and
\figref{figure:stationarity}(b) depict the results of the
$\chi^2$ stationarity test for two seizures, which are again
representative of the range of observed stationarity
phenomena.  These figures utilize the same format as the
variance evolution figures except that black and white
pixels are now used to indicate whether the~pdf of a new EEG
epoch was or was not distinct from the accumulated pdf for
previous epochs at the 95\% confidence level.

The $\chi^2$ test results for the seizure depicted in
\figref{fig:global-variance}(b) appear in
\figref{figure:stationarity}(a). While the data in
\figref{fig:global-variance}(b) are relatively nonstationary
by the variance analysis, they are not obviously
nonstationary using the $\chi^2$~measure, so that there are
differences between these measures of stationarity.  In
\figref{fig:global-variance}(a), leads~O2, T3, T5, T4, T6
and~O1 have the longest stationary single-lead segments
during a seizure lasting about 16-40~seconds, which is
consistent with these leads not demonstrating an increase in
variance over the seizure.  Regions of global stationarity
as identified by the $\chi^2$ test were found in all
seizures, with the shortest global segments observed in two
unilateral seizures (e.g., see
\figref{figure:stationarity}(b) where no stationary segments
were observed longer than a few seconds for all leads).  The
range in length of the stationary segments over all leads
was~4 to~30 seconds. All of the seizures had single-lead
stationarity segments of at least 10~seconds, with the
longest single-lead stationary segment lasting 70~seconds.
Fifteen second or larger segments of multi-lead stationarity
were found in half of the seizures.

\subsubsection{Average Power Spectral Frequency}

We found that the~pdf nonstationarity tests and regions of
uniform frequency generally agreed with one
another. \figref{fig:average-frequency}a displays the
average frequency over time for the same seizure depicted in
\figref{fig:global-variance}a and indicates a 10-second
region of nearly constant multi-lead frequency from
10-20~seconds. In contrast, \figref{fig:average-frequency}b
has single leads with long times of stationary average
frequency but this is not seen across all of the leads.
Leads~FP1 and~O2 have greater variation in frequency across
the seizure and have, in general, higher average frequency
content.  The average frequency for both of these seizures
decreases over the seizure. However, in
\figref{fig:average-frequency}b, this decrease does not seem
to occur for leads~FP1 and~O2.  In fact, for seven of the
ten seizures, at least one lead was delayed in decreasing
frequency or did not manifest a decrease in frequency over
the seizure.  This was most commonly seen in the temporal,
occipital, and frontopolar leads.  Across all of the
seizures, fifteen second or longer regions of uniform
frequency in all leads were found in six of the ten
seizures.  Five of ten seizures had many brief global
average frequency nonstationarities and these corresponded
to nonstationarities detected by the~pdf nonstationarity
tests, e.g., time~21 seconds in
\figref{fig:average-frequency}a.

%

\subsubsection*{Summary Of Stationarity Analysis}

A substantial variability was found in the length of
stationary segments across the seizures studied.  For most
of the seizures, there were substantial windows of
approximately stationary behavior observed in all leads
simultaneously, using several different criteria.
Significant differences between the measures were found
suggesting that stationarity according to one of the
criteria does not insure stationarity by another criteria.
Some of the seizures studied did not have multi-lead
stationarity segments longer than a few seconds, which
suggests that there are likely to be problems when applying
signal analysis techniques that assume signal stationarity
for generalized tonic-clonic seizure data.  On the
other-hand, since these analyses indicate that stationary
segments exist for the vast majority of these seizures, such
analytic techniques may be validly applied after first
verifying that they are being applied to a segment that is
stationary according a range of stationarity tests.  For
both unilateral and bilateral seizures, these analyses also
indicate that there may be leads that are delayed in
demonstrating statistical changes such as an increase in
signal variance or a decrease in average frequency as
compared with other leads.  This was seen most commonly in
temporal, occipital, and prefrontal regions.

  \subsection{Redundancy}
  \label{results-redundancy}
  \subsubsection{Time Evolution Of Mutual Information Transmission Coefficient}

The time evolution of mutual information for each lead were
calculated with respect to the lead~CZ.  (As discussed
above, CZ was picked because of prior reports suggesting
that ECT-induced seizures tend to be maximal in amplitude in
this region~\cite{Weiner91}. However we verified that the
results were the same when mutual information was studied
with respect to a number of different leads.)  The time
evolution of mutual information over the course of two
seizures that illustrate the range of phenomena seen in the
10~seizures are depicted in \figref{figure:time-mutual}a and
\figref{figure:time-mutual}b.  As in previous figures, the
amount of mutual information shared between each lead and~CZ
is represented by color pixels whose magnitude is indicated
in the color bar below each figure.

The ten seizures varied substantially in the spatiotemporal
pattern of redundancy as measured with mutual information.
\figref{figure:time-mutual}a illustrates a seizure with high
redundancy of CZ with most other leads, whereas
\figref{figure:time-mutual}b depicts a seizure with little
inter-lead redundancy with CZ, even though this seizure has
a large stationary region.  \figref{figure:time-mutual}a
also demonstrates that over the course of the seizure there
is a transition from a period of relatively low-interlead
redundancy to increased redundancy where the greatest
inter-lead redundancy, like the greatest variance, is in the
fronto-central region.  The increase in redundancy and
amplitude, although not manifest in all seizures, coincides
with the transition from the tonic to clonic phases of the
seizures~\cite{Weiner91}.  In addition, as with variance and
average frequency, some leads are late to increase in
redundancy or do not do so at all.  This was true for 5 of
the 10 seizures and most often occurred in the temporal and
occipital leads.

The mutual information also provided information about the
redundancy of stationarity for the EEG data.  There was a
period of uniform redundancy for~6 of the~10 seizures
ranging from~10 to~20 seconds in length.  These regions were
located in the mid-ictal region of the seizure, coinciding
with the clonic phase. The regions of constant redundancy
across the leads coincided with regions where~CZ was
stationary and also where the frequency content of all the
leads was nearly uniform.

\subsubsection{Correlation Coefficient}

The amount of redundancy among seizures was also studied by
calculating the time-evolution of the correlation
coefficient of each lead with CZ.  Two seizures illustrating
the range of patterns of inter-lead correlation among the
ten seizures appear in \figref{figure:time-correlation}a and
\figref{figure:time-correlation}b.  Similar phenomena are
seen with the correlation coefficient as with mutual
information.  All leads have low redundancy initially
followed by an increase in the mid-ictal period, with the
greatest redundancy among the fronto-central leads with
correlation coefficients ranging from 0.7-0.95.  Some leads
are late to increase or never increase in redundancy.  In
both of the figures, this occurs most prominently for the
fronto-polar, temporal, and occipital leads.  Late onset of
an increase in redundancy occurred in 8 of the 10 seizures
and was most consistent and most pronounced in the occipital
and temporal leads.  In terms of the stationarity of
redundancy, there were regions of uniform correlation for~5
of~10 seizures ranging from~10 to 20~seconds in length.  As
discussed below, one factor that must be considered for
accounting for leads with decreased redundancy is the
possibility of a phase lag between the signals in the leads
studied.

\subsubsection{Average Mutual Information For Stationary
Midictal Segments}

Average mutual information calculations were performed on
regions that were found to be stationary as determined by
the $\chi^{2}$ nonstationarity test.  Stationary regions
were identified that were 6 to 20 seconds in length within
the mid-ictal portions of each seizure.  The average mutual
information for two seizures, that are representative of the
phenomena we observed among the ten seizures, is depicted in
\figref{figure:ave-mutual}a and \figref{figure:ave-mutual}b.
In these figures, the darkness of the square at the
intersection of two lead labels indicates the mutual
information shared by those leads.  The correspondence
between the degree of shading and the degree of redundancy
is indicated by the scale at the left of the figure. Note
that the same information is portrayed in the upper left and
lower right halves of these figures.

Once again we found that the greatest interlead redundancy
tended to occur in the frontocentral regions.  The regions
of highest redundancy tended to differ for~UL- (see
\figref{figure:ave-mutual}a) and~BL-induced ECT seizures
(see \figref{figure:ave-mutual}b).  There tended to be
increased redundancy in the right (stimulated) hemisphere
for UL~ECT (note the clustering of lighter boxes in the
lower left and upper right corners of
\figref{figure:ave-mutual}a), whereas for BL~ECT the
redundancy was not localized to either hemisphere.  The
regions of lowest redundancy were the occipital and left
temporal leads in all unilateral seizures.  In bilateral
seizures, all temporal and occipital leads had lowered
redundancy with no hemisphere dependence.

\subsubsection{Inter-Lead Correlation For Stationary Midictal Segments}

The same global stationary regions utilized for average
mutual information analysis were also used for the average
correlation coefficient calculations.  Two representative
seizures are depicted in \figref{figure:ave-correlation}a
and~\figref{figure:ave-correlation}b.  Relatively weaker
correlations of the frontal-polar and occipital leads (see
the bands of darkly shaded squares associated with these
leads in \figref{figure:ave-correlation}a and
\figref{figure:ave-correlation}b) with other leads on the
scalp were found in seven of ten seizures with the frontal
polar regions having the poorest correlations.  In some
cases, these leads have poor correlations due to the
presence of large time-delays with other leads on the head
(see below).  Lowered redundancy for the frontopolar leads
in \figref{figure:ave-correlation}b was not indicated by the
mutual information measure for the same seizure
\figref{figure:ave-mutual}b since mutual information was
less sensitive to time delays.  A reduced redundancy was
thus found particularly for the frontopolar and occipital
leads in the midictal period.  In some instances, this was
due to time delays but, in several other cases, the data in
the leads were not as well related.

  \subsection{Interlead Time Delays}
  \label{time-delays}
  A number of leads had sustained poor redundancy with~CZ
because of time delays with~CZ, which artificially lowered
the apparent redundancy in our correlation analysis.  This
is seen in~6 of~10 seizures in leads FP1, FP2, O1, and O2
(see Fig.~\ref{figure:time-correlation}a).  In contrast,
\figref{figure:time-correlation}b illustrates an instance
where leads~O1, O2, FP1, and~FP2 have poor correlations
with~CZ for reasons other than time delays.  Because of the
dynamical and physiological importance of consistent time
delays across the head in the midictal period, we sought to
understand this phenomenon more thoroughly as we now
discuss.

The time delay calculation cannot be calculated for data
where the leads have a low redundancy since if two leads are
not related, then their time delay has no meaning.  As a
result, we required that segments meet the
$\chi^2$-stationarity criterion, be of constant average
frequency by visual inspection, and have interlead
redundancy of at least~0.5 as measured by mutual
information. We first studied time delay over the course of
the seizures.  Segments of uniform nonzero time-delay over
many leads were identified in~7 of the~10 seizures studied
and they lasted between~4 and 20~seconds.  The most
consistent pattern in time delays was a large delay from the
front to the back of the head.  This pattern was found in~4
of the~10 seizures, and the magnitude of the delay ranged
from -10 to 15~ms. The occipital leads had a negative time
delay with CZ whereas the prefrontal leads had a positive
time delay.

The average amplitude of the time delay with respect to CZ
across the head for one representative mid-ictal segment is
mapped in \figref{figure:time-delay}.  This map depicts the
time delay with respect to CZ for the 19 scalp leads such
that the color at each point indicates the degree of time
delay based on the scale at the bottom of the figure. Data
is mapped at the 19 white marks corresponding to lead
location with linear interpolation in between leads.  This
figure illustrates what appears to be a consistent
frontal-to-occipital time delay indicative of wave-like
propagation from the occipital to prefrontal regions during
the mid-ictal portion of some seizures.  In some instances
we observed a tendency for counter-clockwise rotations
around CZ.
%
%
While this wave phenomena was mapped for a brief period, it
could be observed for over~20 seconds.

\section{Conclusions}
\label{conclusions}



In this paper, we have studied the stationarity and
redundancy of multielectrode EEG data during GTC seizures.
Stationarity of EEG data is a main concern in gaining
insights to the underlying dynamics.  Reliable statistical
analysis and many nonlinear measures of complexity assume
statistically stationary data.  We find that the use of
variance non-stationarity tests, pdf non-stationarity tests,
and average-frequency evolution may provide a useful measure
of the stationarity in these signals. Global regions of
stationary data are of length 8 to 20 seconds in the
majority of seizures and single lead stationary regions are
typically of length 20 to 40 seconds.  Highly non-stationary
seizures tend to have higher variance and show no global
regions of stationarity. In prior work, we have found that
seizures which had a more predictable EEG pattern over time
(according to the largest Lyapunov exponent) were more
therapeutically effective~\cite{Krystal97}.  Further work
would be useful to establish whether the more stationary
seizures are also more beneficial in the treatment of
depression.

Redundancy on the surface of the head varies among patients
and is further complicated by nonstationarities and leads
that are late or never enter into the seizure activity.  We
find that the mutual information statistic is much more
robust to time-delays between different leads than linear
correlation measures.  The frontal region of most seizures
along the excitatory current path tends to have sustained
higher redundancies than other portions of the head.  More
interesting is the high redundancy seen in some seizures
among spatially distant portions of the head.  In the
majority of seizures, the frontal and occipital regions are
poorly redundant. However, in three seizures we find high
redundancy between different pairings of anterior and
posterior leads.  High redundancy pairings, O2-FP1 and
O2-F8, seen in some seizures are not understood.  In all
seizures, the redundancy tends to increase in the later
mid-ictal portions of these seizures.  This also coincides
with seizures exhibiting more rhythmicity and spatial wave
behavior.

The wave-like behavior in the EEG which we have observed in
GTC seizures has not been previously reported.  This
behavior could be caused collectively by the coupling of
dynamically different parts of the brain, leading to a wave
that propagates cyclically through the brain tissue.
Alternatively, one region of the brain may be a source that
drives the observed seizure activity in other parts of the
brain. Unfortunately, physiologic evidence is presently
lacking that can determine the mechanism of this wave-like
electrical activity pattern.  Identification and
characterization of such cortical waves may be possible by
using a complex Karhunen-Lo\`eve
decomposition~\cite{Horel84}, which has been used
successfully by meteorologists to identify wave motion in
the atmosphere. Such analysis may help to determine how
frequently wave motion occurs in GTC seizures and to explain
the mechanism of electrical propagation of activity in these
seizures.

The occurrence of surface waves on the cortex of the head
and the fact that many seizures have leads that are late or
that never enter into the seizure evolution call into
question the utility of the term ``generalized'' in
describing GTC seizures. There is evidence that sometimes
the seizure spreads from the frontal region to the occipital
and temporal regions of the head.  Also, some leads can be
as late as 10~seconds to enter into the seizure or do not
participate in the seizure.  Late-to-generalize seizures
have been previously reported~\cite{Brumback82}, but leads
not involved in the seizure have not previously been
described and the clinical relevance of these uninvolved
leads remains to be established.

These findings speak against the view that GTC seizures are
an instantaneous all-out response of the brain.  Instead
they are consistent with prior work indicating that GTC
seizures are graded rather than maximal
responses~\cite{Niedermeyer93}. Such work involved the
demonstration that the cerebellum and the lower brainstem
had varying degrees of electrophysiological involvement in
GTC seizures in the cat and that the visual evoked response
was variably disrupted in GTC seizures in
humans~\cite{Rodin66}.  The findings of the present study
demonstrate that such variability is manifest in the EEG
signals recorded during GTC seizures as well.  In addition,
there is evidence that there are complex spatiotemporal
dynamics involved in the development and propagation of
these seizures.  Such results speak against the sudden onset
of massive discharge in reticular structures that was once
proposed~\cite{Gastaut72} and more for a graded, diffusive
model for the origin and spread of GTC seizure activity.

In summary, our analysis of the stationarity and redundancy
of multichannel EEG data recorded during GTC seizures has
identified numerous new features and these have implications
for further research.  First, there is a substantial
variability in stationarity.  Techniques that assume
stationarity should be applied only after verifying that a
given time segment is acceptably stationary.  The
variability in stationarity suggests a need for further
studies that could determine the relationship of the degree
of nonstationarity of GTC seizures to their antidepressant
efficacy.  We also found a variation in redundancy between
the leads, with the greatest redundancy in fronto-central
regions with decreased pre-frontal, temporal and occipital
redundancy.  Further work to determine the physiology
underlying this differential spatial redundancy will be
important for understanding GTC seizures as will attempts to
determine whether diminished redundancy in particular regions
is associated with diminished therapeutic efficacy or
side-effects of ECT.  These same regions are also apparently
delayed at times in entering the seizures suggesting a more
complex spatio-temporal evolution than previously reported,
which is another feature that will be important for
understanding the physiology and antidepressant efficacy of
these seizures. Complex spatiotemporal dynamics is also
suggested by evidence of wave-like behavior.  All of these
observations point to graded, rather than all or none
physiologic phenomena underlying GTC seizures and will be
important bases for new models of GTC seizures and in better
understanding and improving ECT treatment.

\section*{Acknowledgements}
This work was supported by grants NSF-CDA-91-23483 and
NSF-DMS-93-07893 of the National Science Foundation, by
grant DOE-DE-FG05-94ER25214 of the Department of Energy, by
grants K20MH01151 and R29MH57532 of the National Institute
of Mental Health, and by a Computational Science Graduate
Fellowship Program of the Department of Energy.



\newpage

\begin{figure}     
\caption{Ten seconds of 19~channel ictal EEG data recorded
during the middle portion of two ECT seizures.  The vertical
lines indicate one second intervals, the vertical axis is
voltage in microvolts.  {\bf (a)} A representative seizure
that has a more stationary mid-seizure EEG pattern and
higher redundancy among the leads.  {\bf (b)} A
representative seizure that has a relatively lower
stationarity and interlead redundancy.  }
\label{fig:ictal-eeg-data}
\end{figure}

\begin{figure}    
\caption{The location on the head of the 19~EEG leads
according to the International 10-20 System. These are
displayed in a two-dimensional representation looking down
on top of the head. }
\label{fig:electrode-location}
\end{figure}

\begin{figure}    
\caption{Variance $\sigma^2(t)$ versus time~$t$ for
seizures~A and~B. The period of postictal suppression
following the end of the seizures is excluded from all
figures and occurred just following the period depicted in
the figures. (Epochs are in units of 2 secs).  (a) Variance
evolution for BL~ECT seizure~A with a 15-second stationary
segment from $t=0$ to $t=15$.  The highest variance is seen
in~FZ and~PZ.  Leads~T6, T4, O2, O1, and~T5 have lower
variances and are late to enter the seizure. {\bf (b)} The
variance evolution for BL seizure~B shows both an initial
stationary region and global regions of nonstationarity at
epochs 13, 19, and~21. }
\label{fig:global-variance}
\end{figure}

\begin{figure} 
\caption{ {\bf (a)} Nonstationarity pdf test for BL~seizure~B,
based on a comparison of probability distribution functions
(pdfs) using a $\chi^2$-test at the 95\% significant level
(see text for details). This is a relatively stationary
seizure.  {\bf (b)} Nonstationarity pdf test for
UL~seizure~B typifies a highly nonstationary seizure with
few regions of global stationarity.}
\label{figure:stationarity}
\end{figure}

\begin{figure} 
\caption{ {\bf (a)} Average frequency $\langle\omega\rangle$
versus epoch for BL~seizure~A. This seizure has a region of
uniform average frequency between 12 and~21 seconds.  {\bf
(b)} Average frequency for unilateral (UL) seizure~A.  
Leads O2 and FP1 never slow down in frequency 
unlike the other leads.}
\label{fig:average-frequency}
\end{figure}

\begin{figure}  
\caption{ {\bf (a)} Mutual information coefficients for
BL~seizure~A demonstrates that (particularly initially)
leads T6, O2, FP2, O1, P3, T5, T3, F8, and T4 have
relatively lower redundancy compared with the other
leads. This suggests that they are late to enter the
seizure.  At 8~epochs, all leads have improved mutual
information.  Lower redundancy is seen in occipital and
frontal-polar leads throughout the seizure. {\bf (b)} Mutual
information coefficients for BL~seizure~B reveals a low
redundancy with~CZ even during the 5-epoch global stationary
region at the beginning of this seizure.}
\label{figure:time-mutual}
\end{figure}

\begin{figure}  
\caption{ {\bf (a)} The linear correlation coefficients between
CZ and the other leads of BL
seizure A clearly shows that the frontal polar leads FP1 and
FP2 are poorly correlated.  This is due to large time-delays
with CZ.  {\bf (b)} The linear correlation coefficients between
CZ and the other leads of
UL~seizure~A demonstrates that channels~O2 and~FP1 are
poorly correlated, a result of their higher frequency
content.  Channels~O1 and~FP2 have poorer correlations since
these leads have large time-delays with CZ.}
\label{figure:time-correlation}
\end{figure}

\begin{figure} 
\caption{ {\bf (a)} Average mutual information transmission
coefficients calculated for UL~seizure~C over a multi-lead
stationary region located by pdf stationarity tests of
8~seconds.  Redundancy is strongest with leads that are in
the same hemisphere of the brain. {\bf (b)} Average mutual
information coefficients for BL~seizure~A over a multi-lead
stationary segment of 16~seconds by the pdf stationarity
test.  There is strong redundancy in the frontal portion of
the brain.  Poor redundancy is seen in the occipital and
temporal leads.}
\label{figure:ave-mutual}
\end{figure}

\begin{figure}  
\caption{ {\bf (a)} Average correlation coefficients of
UL~seizure~A over a multi-lead stationary segment of
16~seconds by the pdf stationarity tests.  Channels~O2
and~FP1 have good correlations with each other but not with
other leads on the head due to their having high average
frequencies.  Channels~FP2 and~O1 have the same common
frequencies with other leads on the head, but have lower
correlations due to significant time-delays with spatially
distant leads.  {\bf (b)} Average correlation coefficients
of BL~seizure~A over a 16~second stationary segment by the pdf
stationarity test and uniform frequency.  The
frontal polar leads have high correlations with the
surrounding frontal leads but poor correlation with the
occipital and temporal leads.  This pattern of time-delay is
due to an anterior to posterior time delay (see
\figref{figure:time-delay}).  }
\label{figure:ave-correlation}
\end{figure}

\begin{figure}  
\caption{ Time-averaged time-delay for BL seizure~A, in the
uniform region (\figref{figure:time-mutual}a ) 
from 25 to 45~seconds . The degree of
average time-delay is referenced according to the color-bar
below the figure. This map, looking down on the top of the
head, portrays 19~data points (represented by the "+"
symbols) with linear interpolation carried out in between
the leads.  Note the significant anterior-to-posterior
time-delay topography.  }
\label{figure:time-delay}
\end{figure}


\newpage
Figure 1a:\\
\ \\
\centerline{\epsfxsize=5in\epsfbox{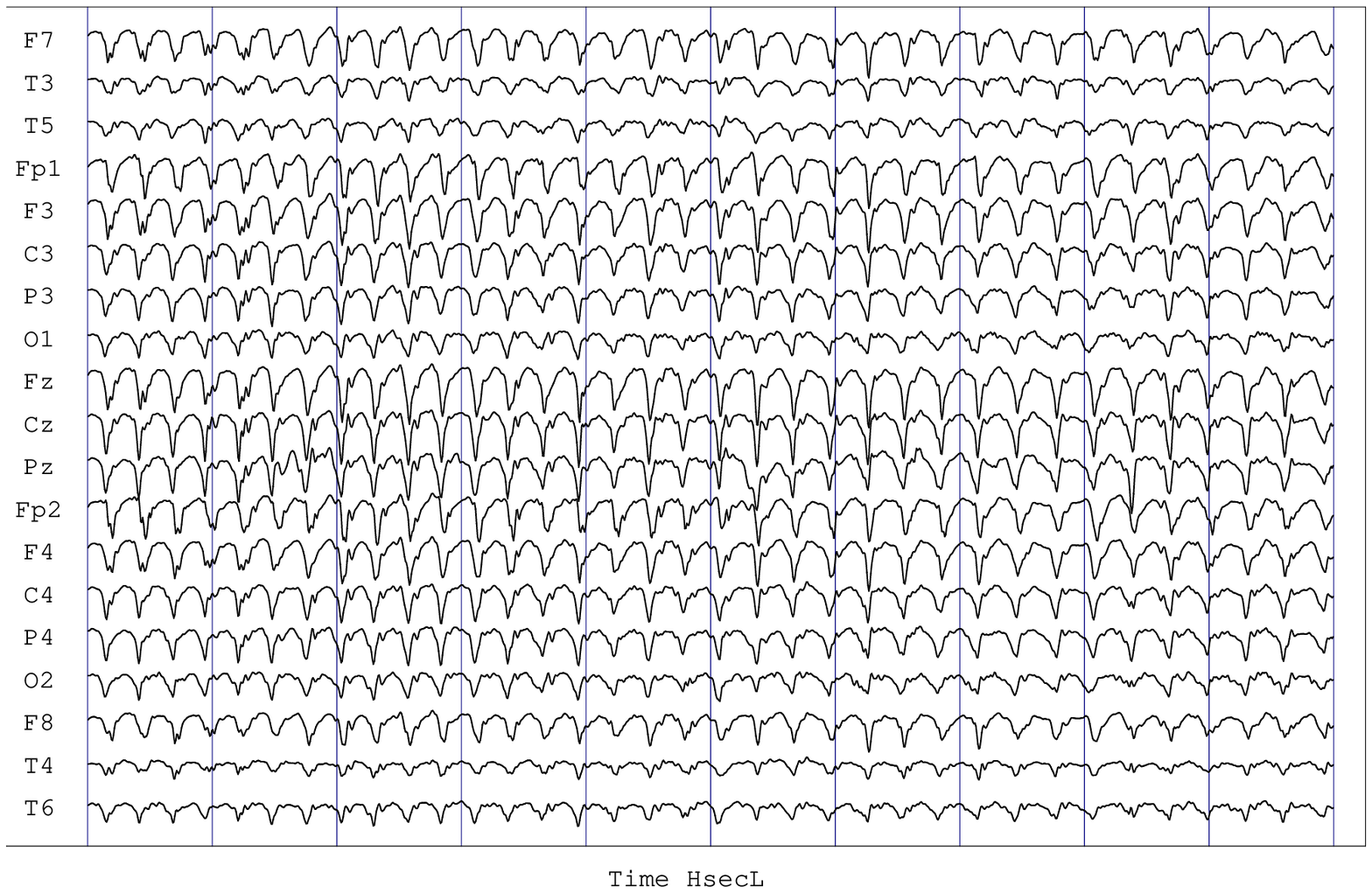}}
\ \\
Figure 1b:\\
\ \\
\centerline{\epsfxsize=5in\epsfbox{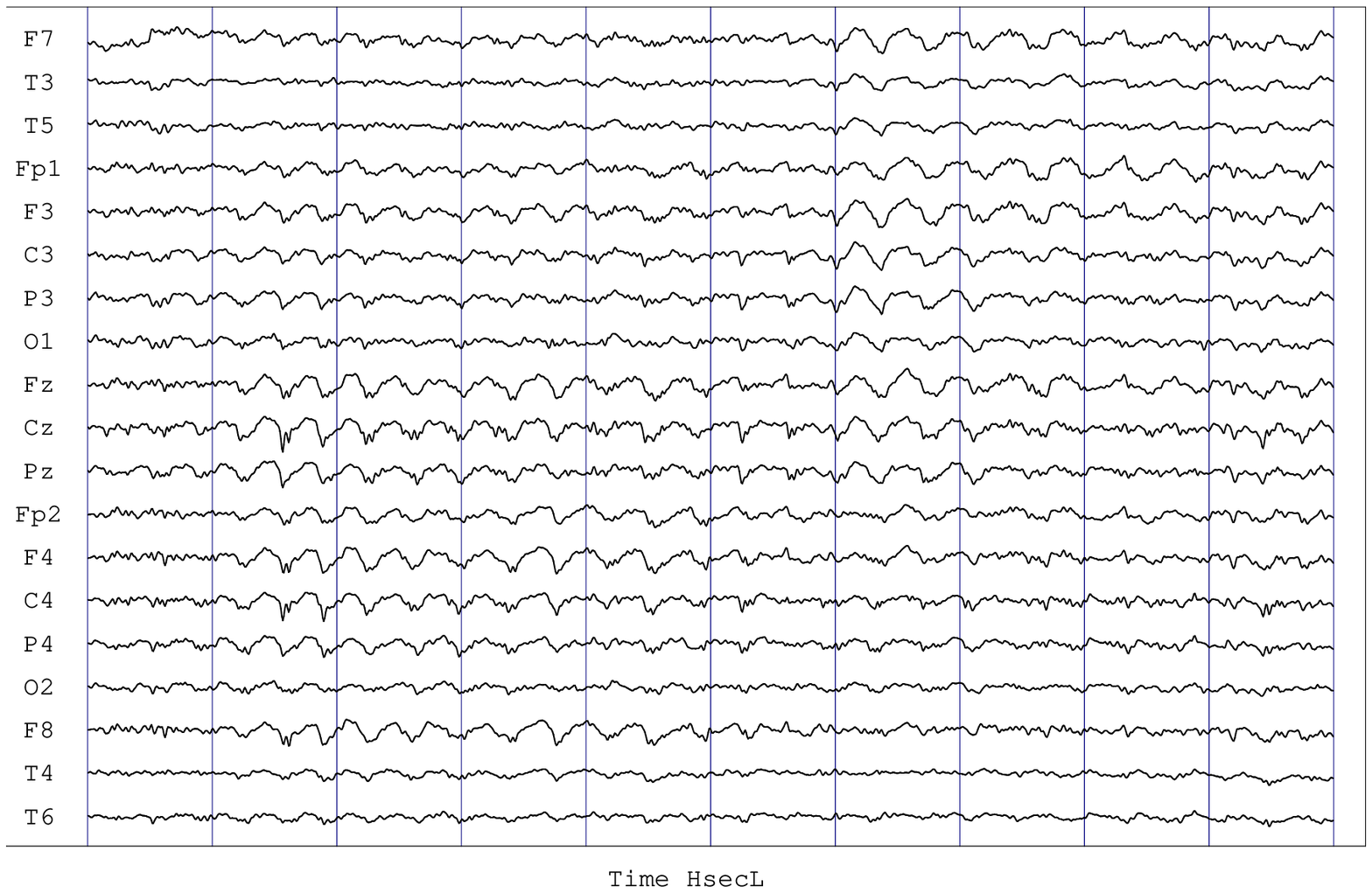}}
\newpage

\newpage
Figure 2:\\
\ \\
\centerline{\epsfxsize=6in\epsfbox{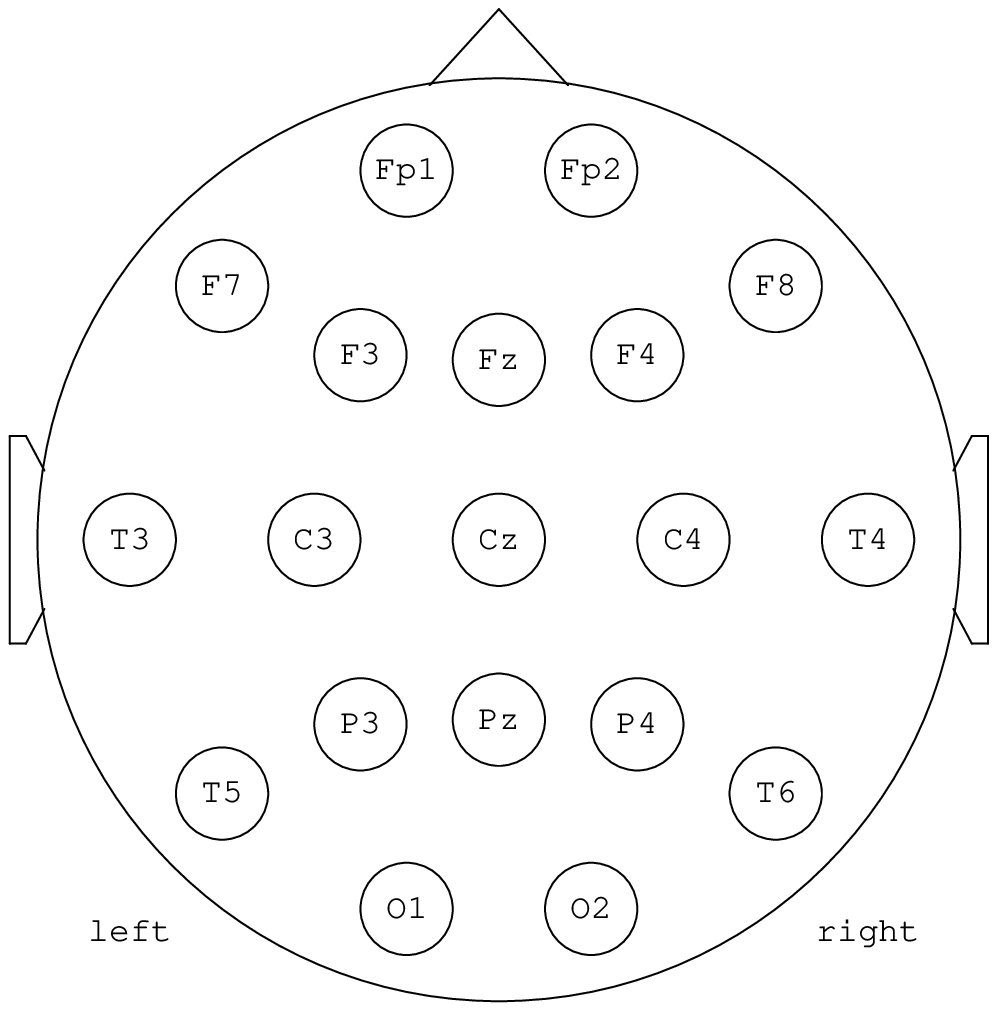}}
\newpage

\newpage
Figure 3a:\\
\centerline{\epsfxsize=6in\epsfbox{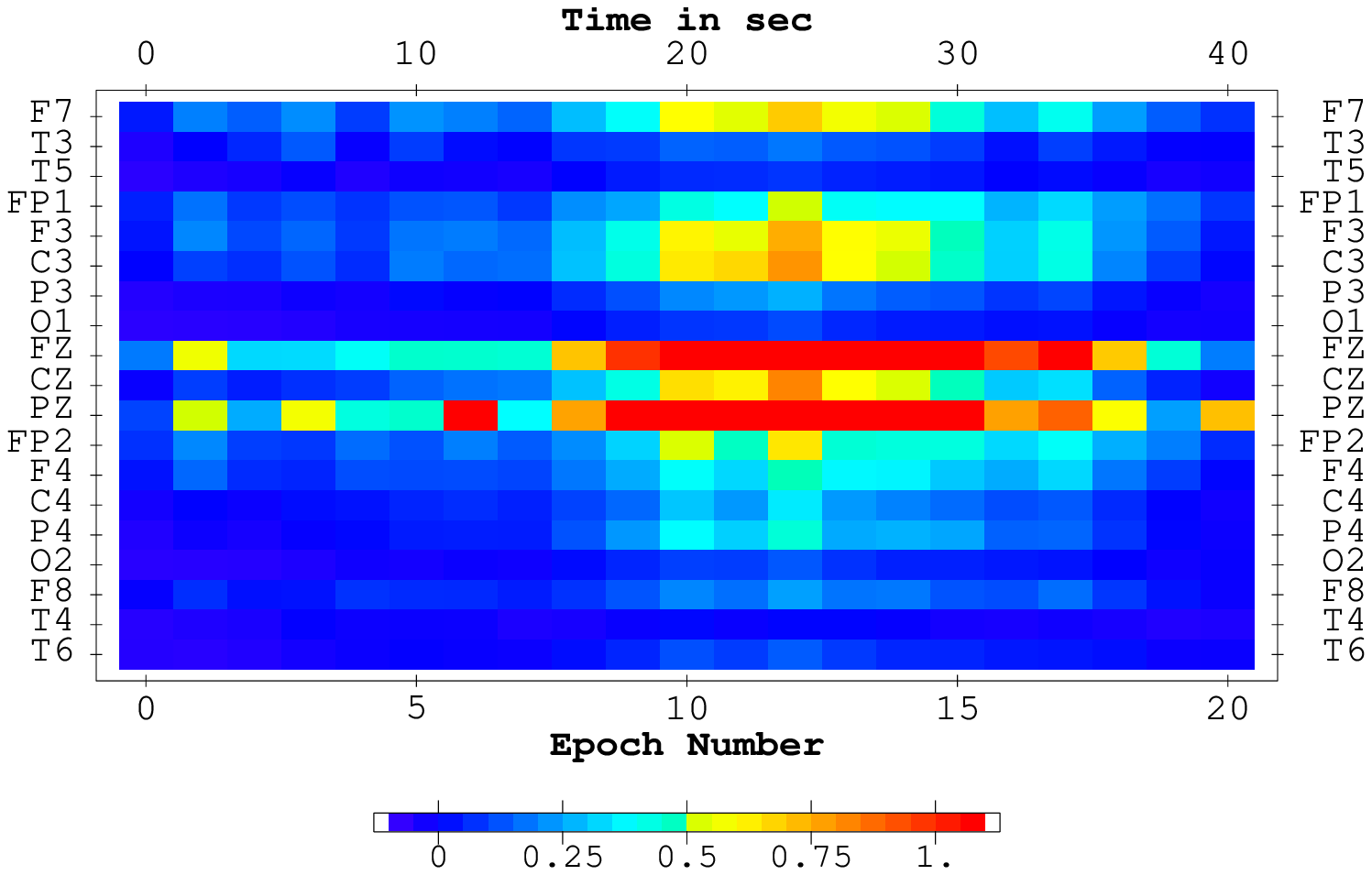}}
Figure 3b:\\
\centerline{\epsfxsize=6in\epsfbox{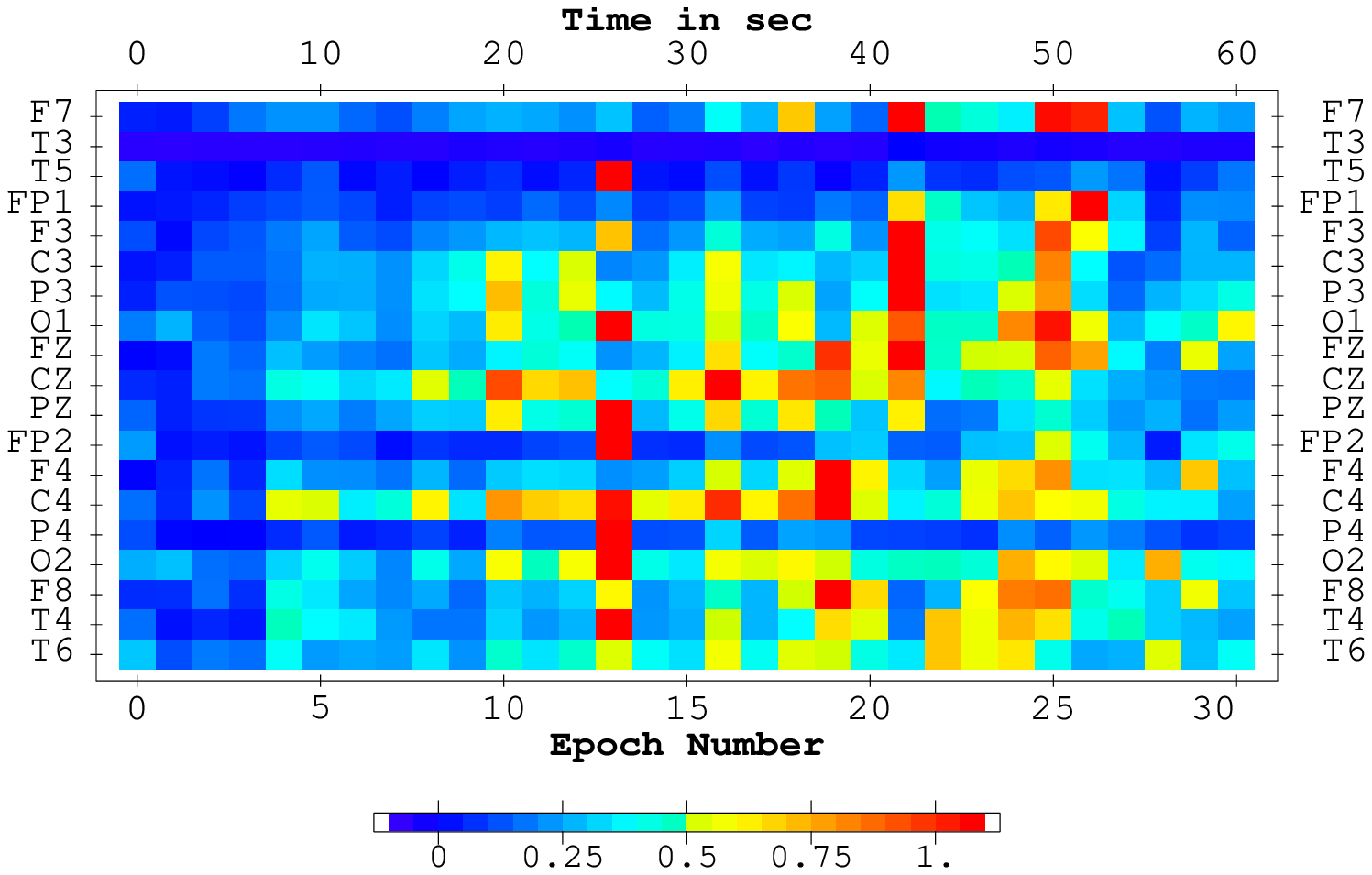}}

\newpage
Figure 4a:\\
\centerline{\epsfxsize=6in\epsfbox{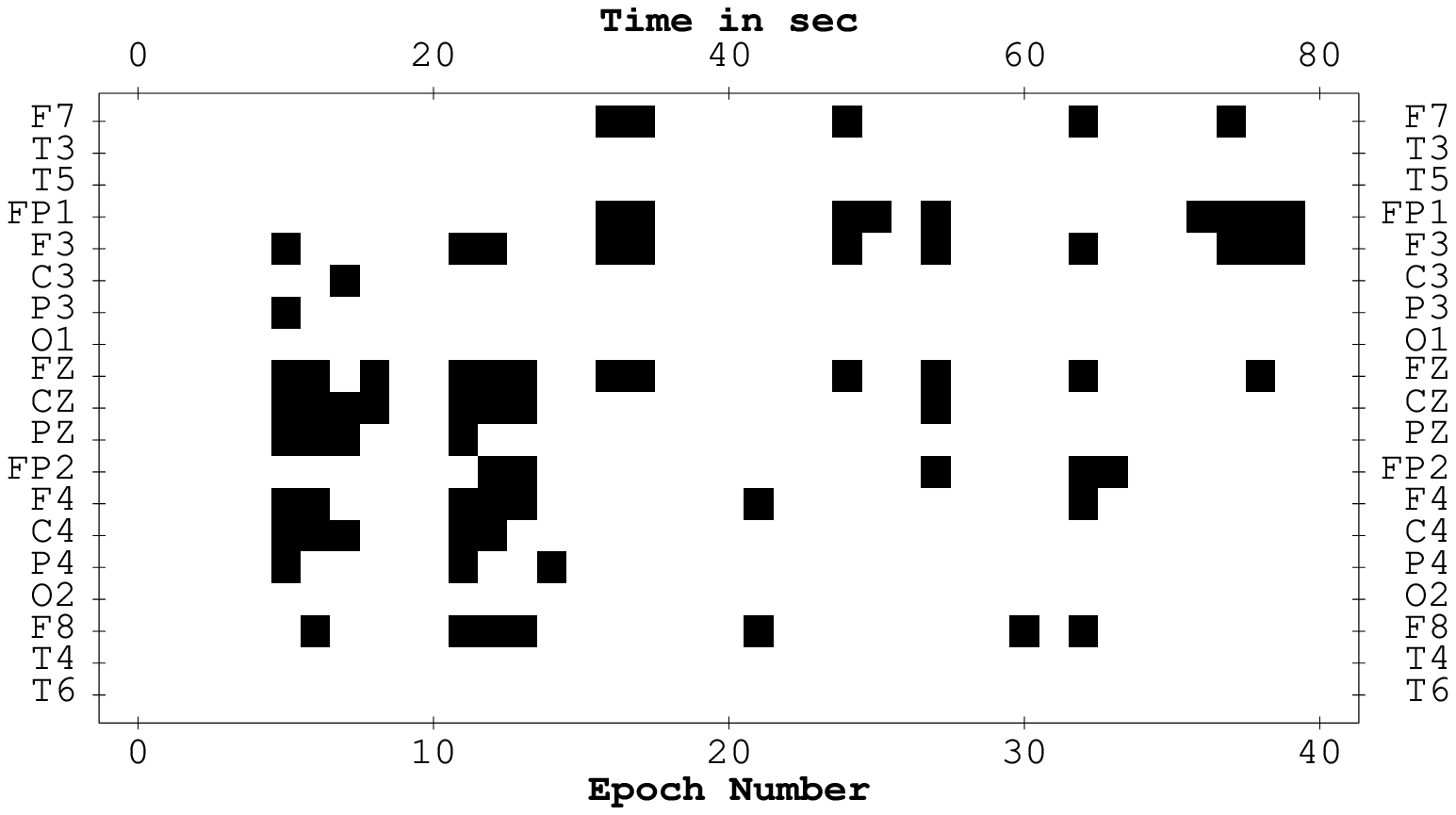}}
Figure 4b:\\
\centerline{\epsfxsize=6in\epsfbox{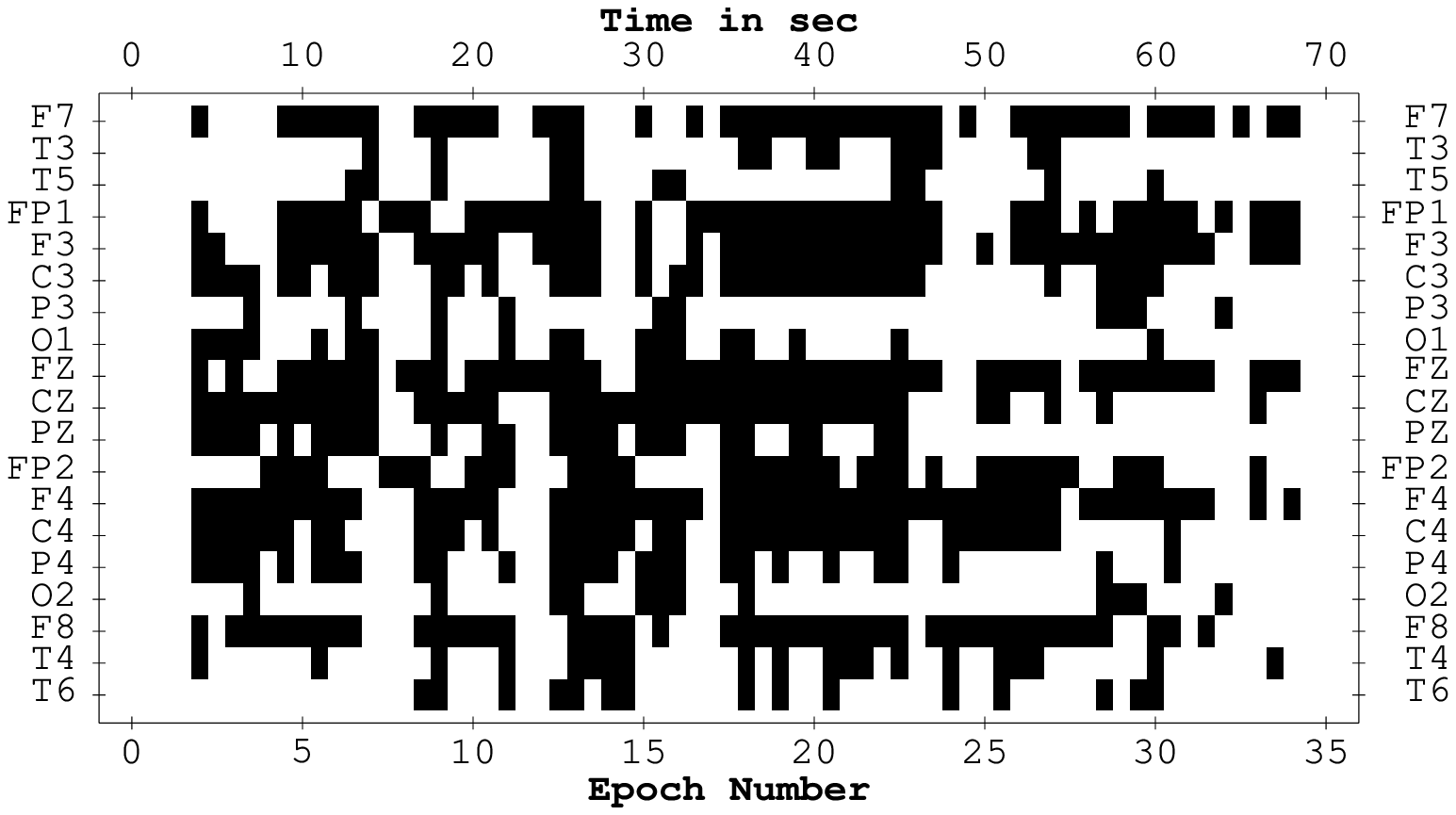}}

\newpage
Figure 5a:\\
\centerline{\epsfxsize=6in\epsfbox{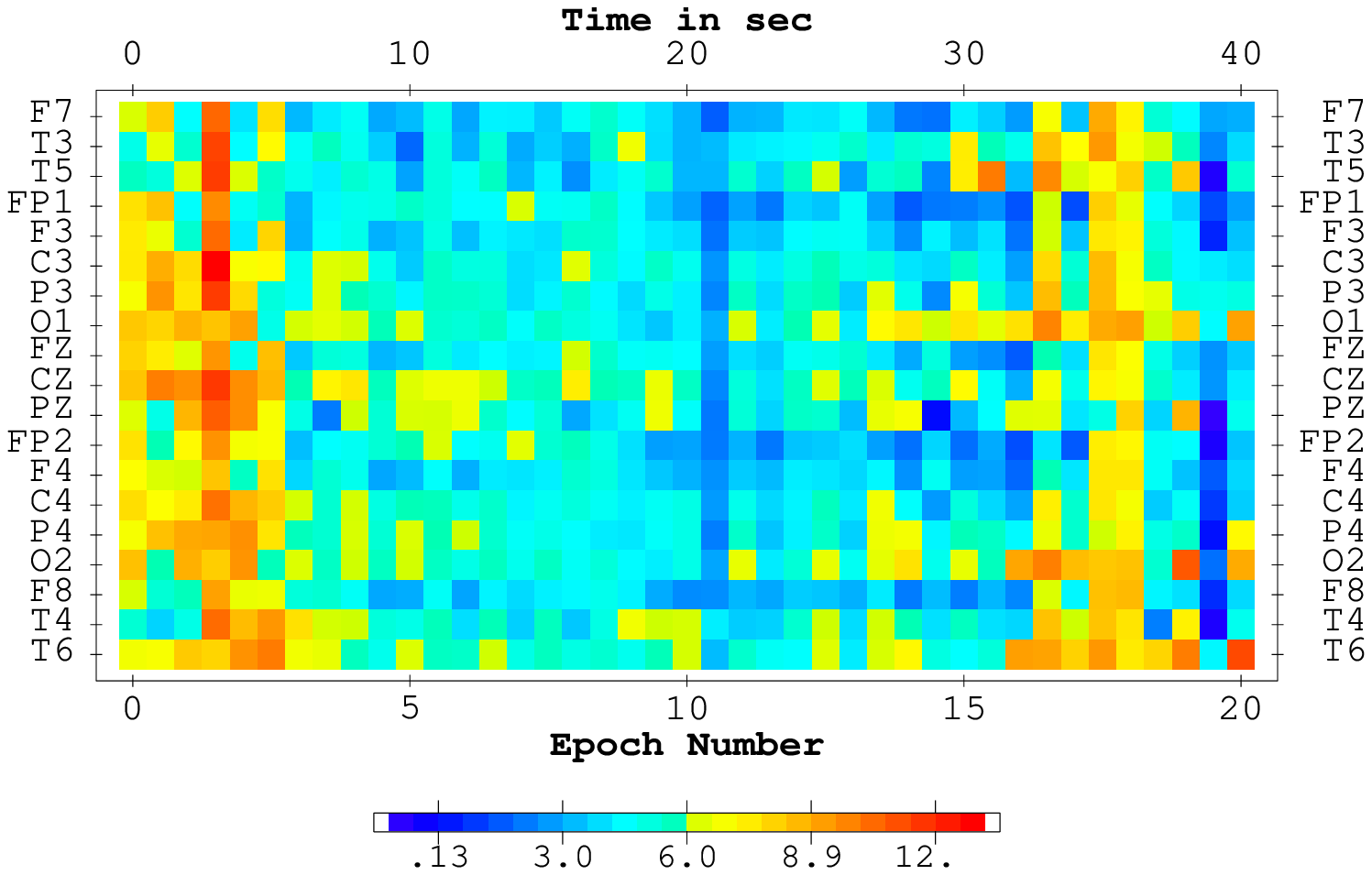}}
Figure 5b:\\
\centerline{\epsfxsize=6in\epsfbox{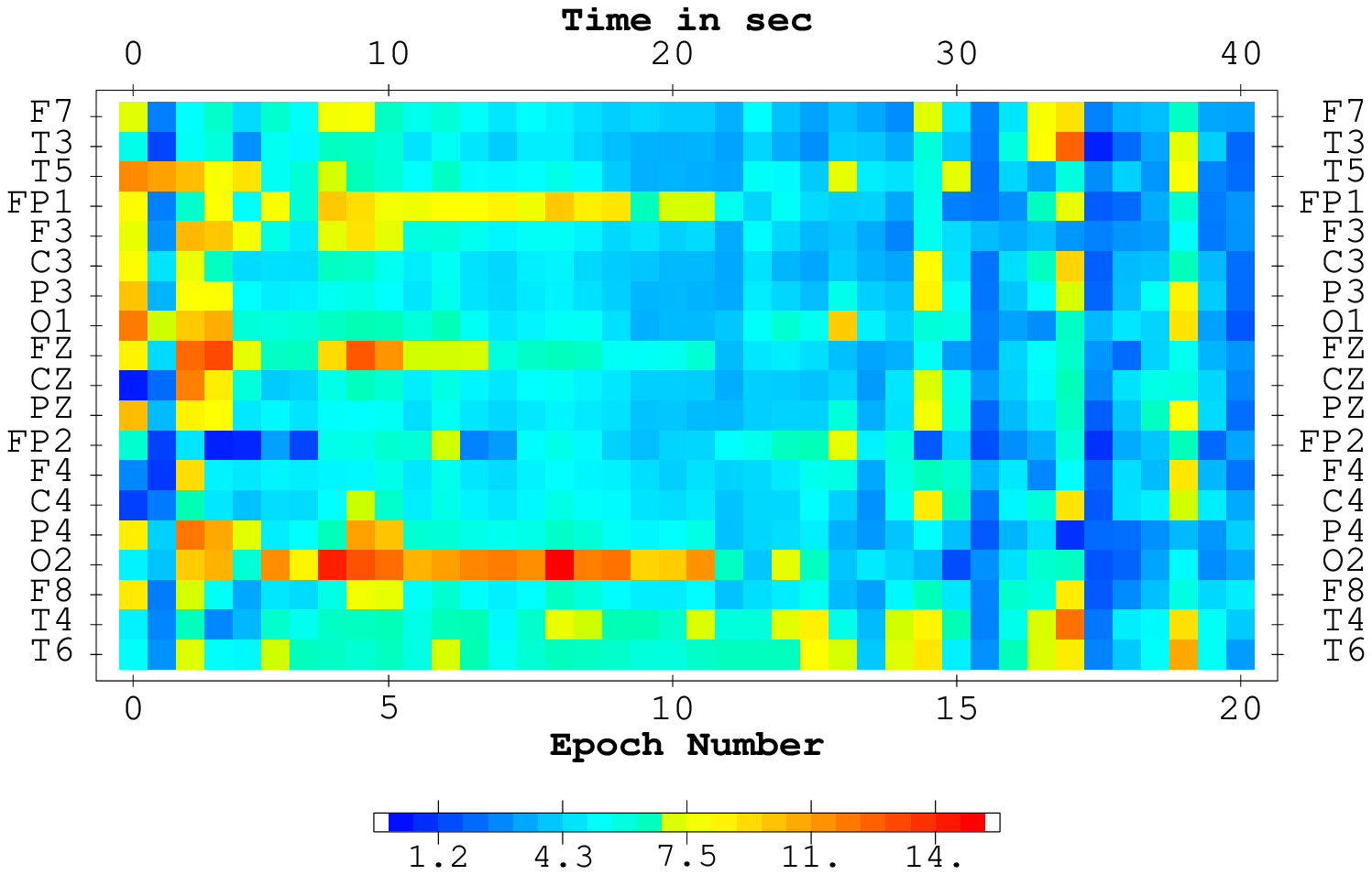}}

\newpage
Figure 6a:\\
\centerline{\epsfxsize=6in\epsfbox{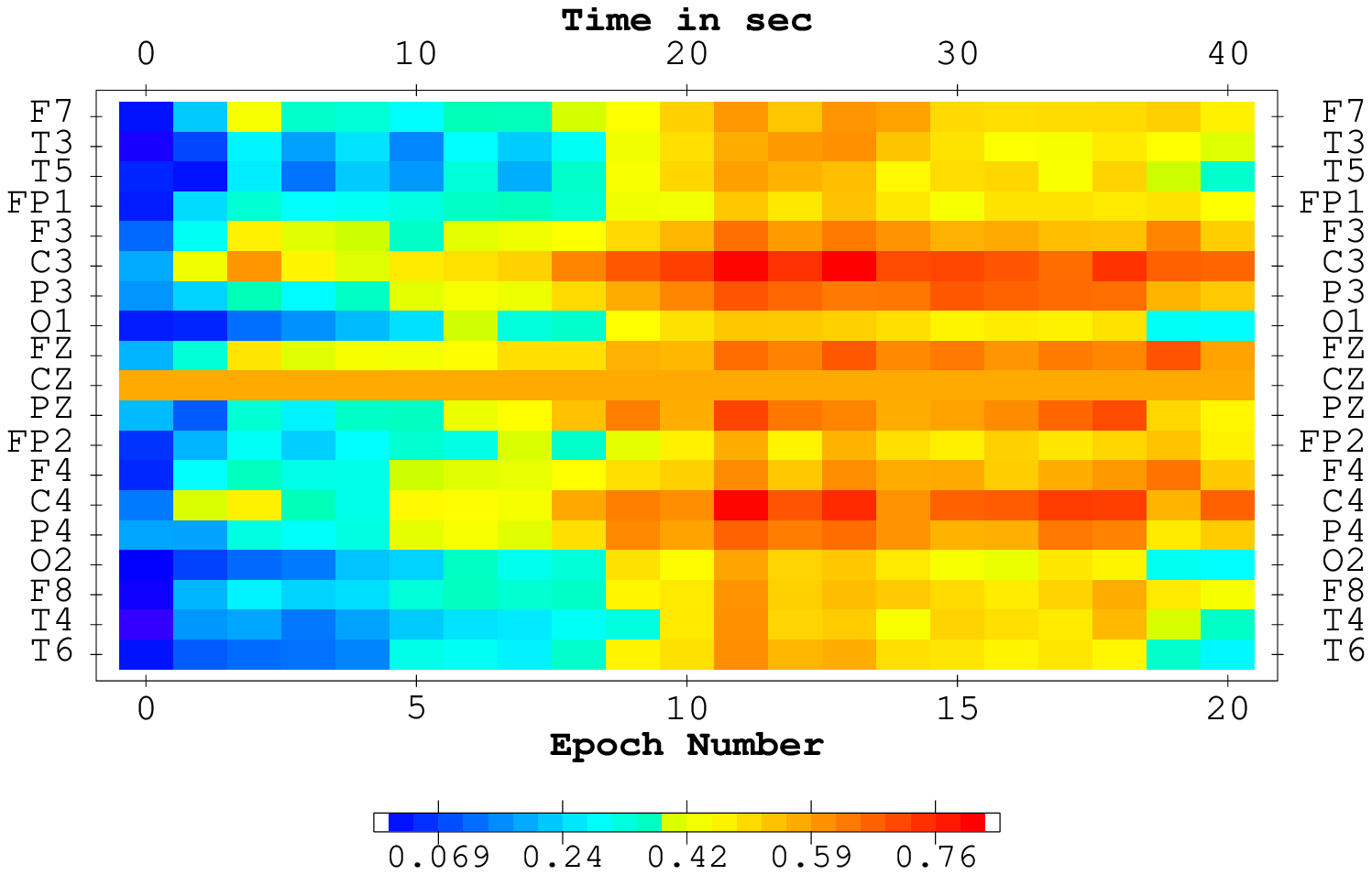}}
Figure 6b:\\
\centerline{\epsfxsize=6in\epsfbox{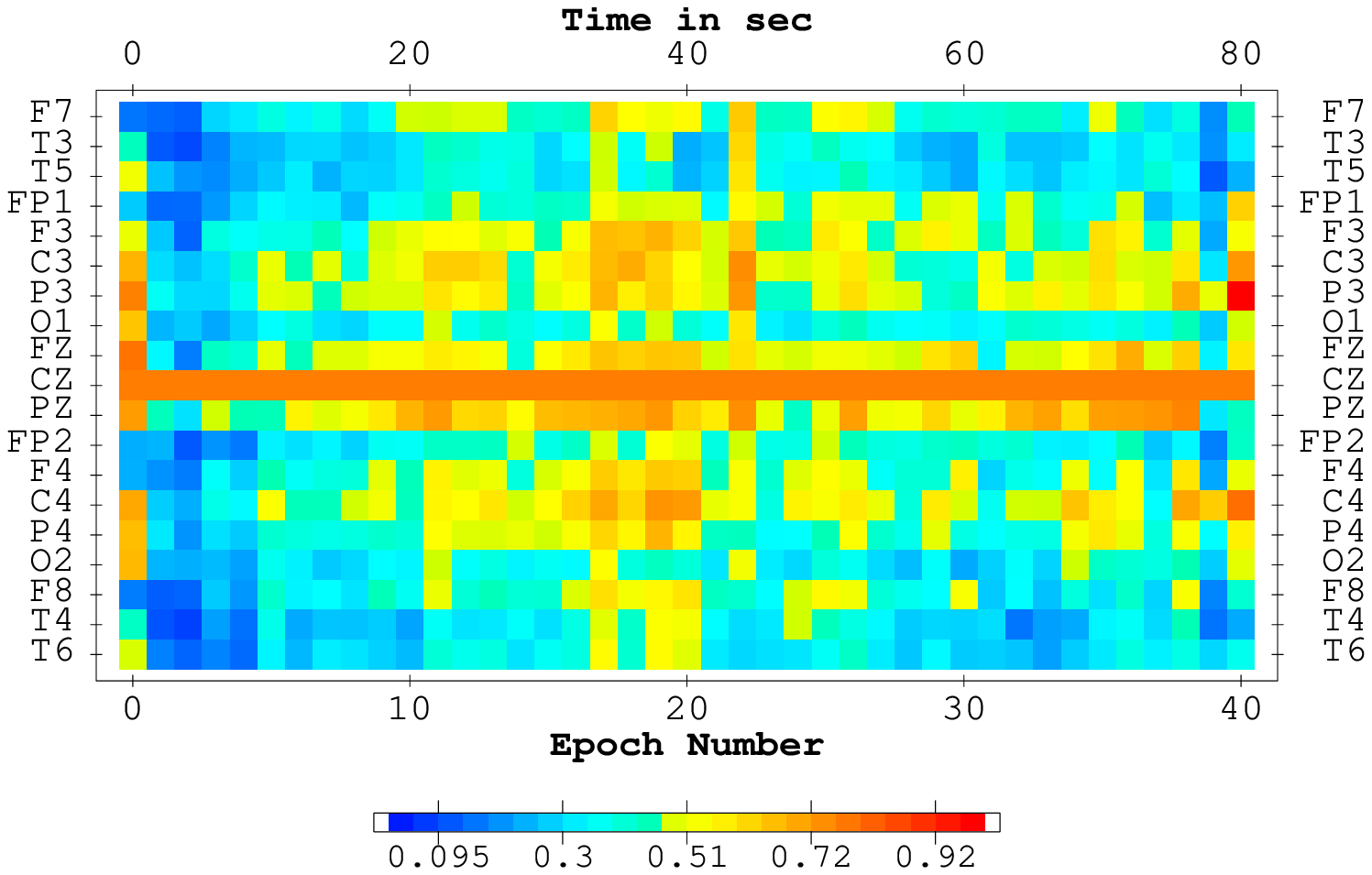}}

\newpage
Figure 7a:\\
\centerline{\epsfxsize=6in\epsfbox{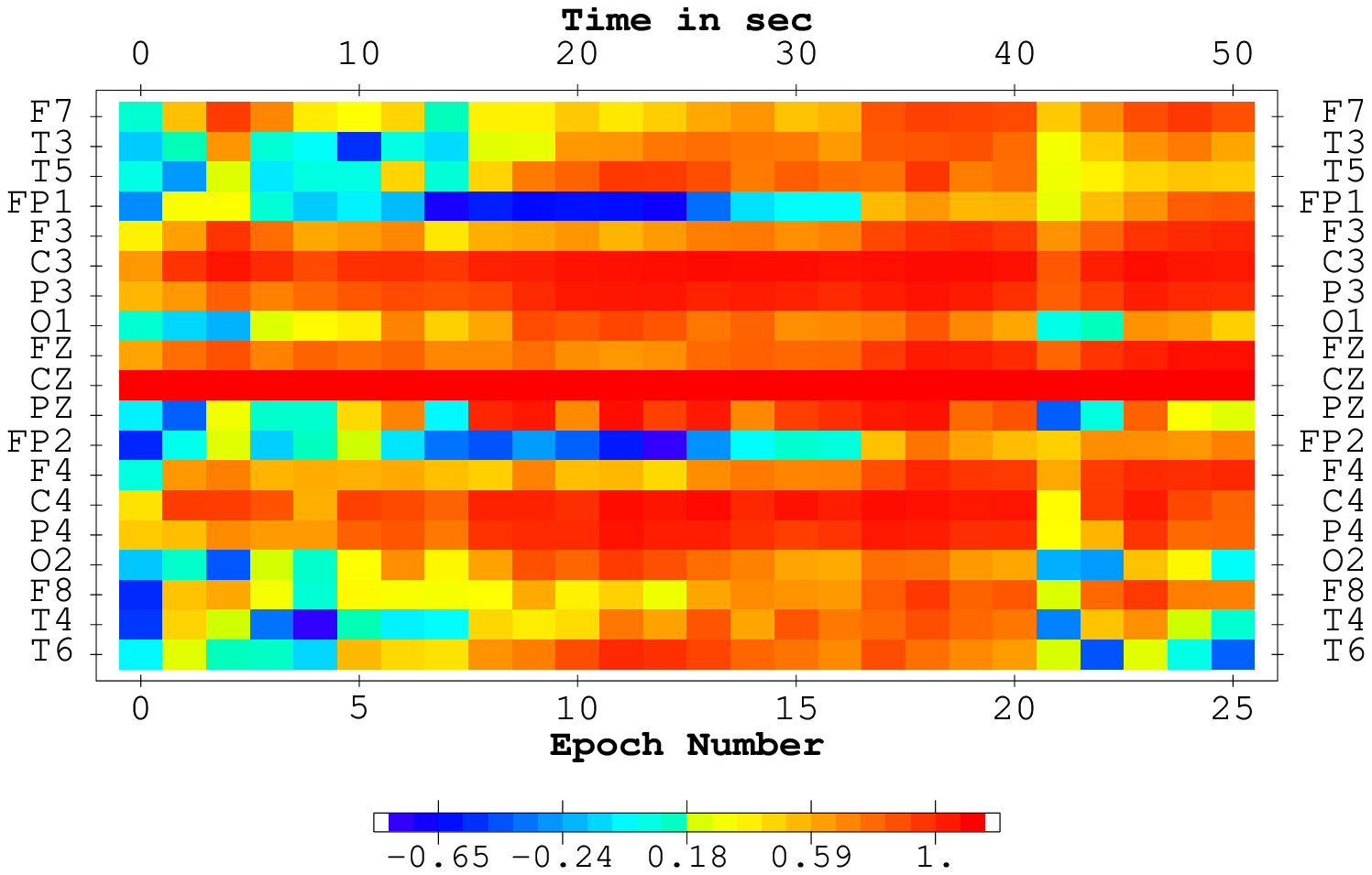}}
Figure 7b:\\
\centerline{\epsfxsize=6in\epsfbox{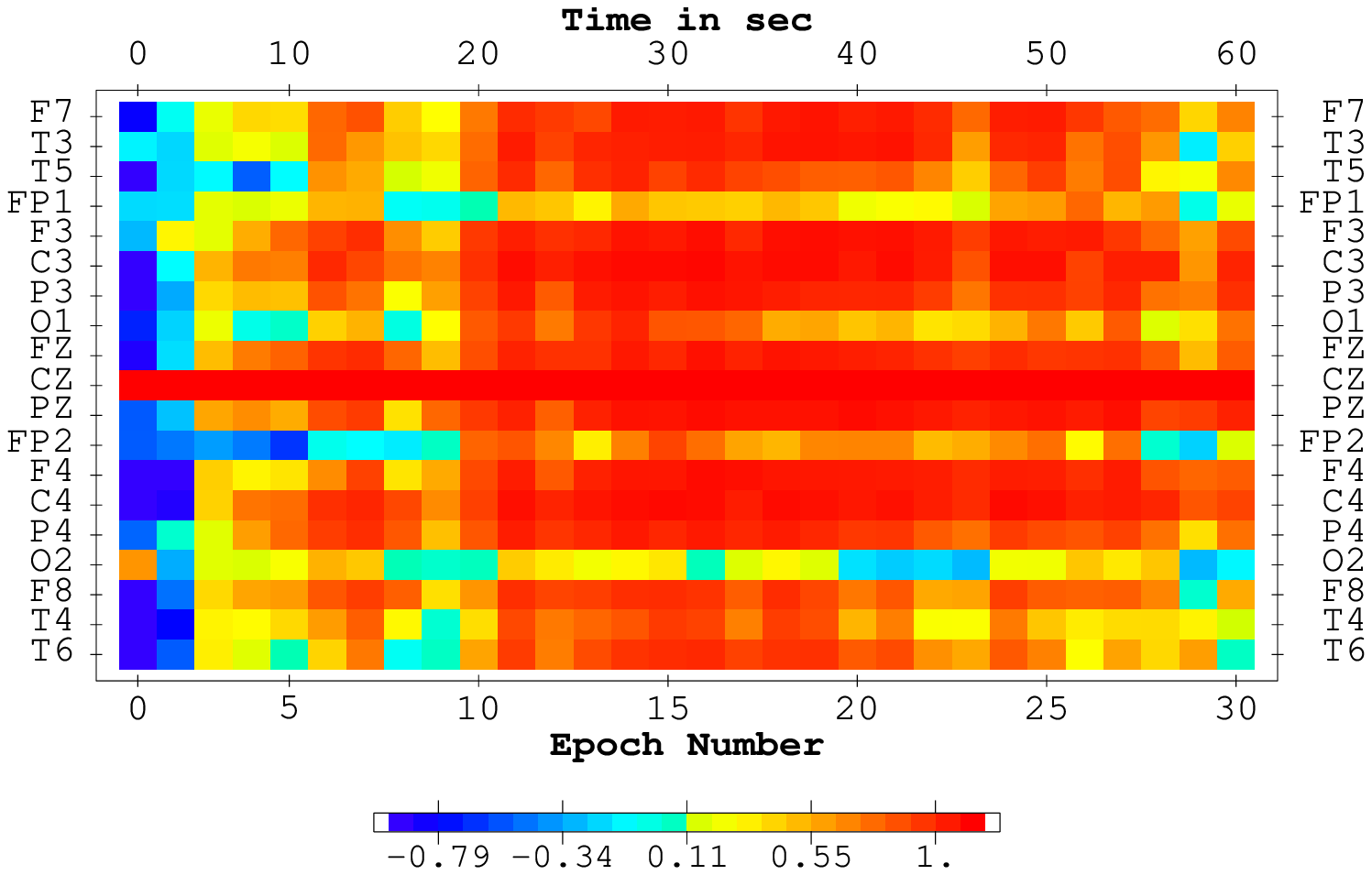}}

\newpage

Figure 8a:\\     
\vspace{.2in}
\centerline{\epsfxsize=5in\epsfbox{Figures/fig8a.epsi}}

Figure 8b:\\     
\vspace{.2in}
\centerline{\epsfxsize=5in\epsfbox{Figures/fig8b.epsi}}

\newpage

Figure 9a:\\  
\vspace{.1in}
\centerline{\epsfxsize=5in\epsfbox{Figures/fig9a.epsi}}

\vspace{.4in}
Figure 9b:\\  
\vspace{.1in}
\centerline{\epsfxsize=5in\epsfbox{Figures/fig9b.epsi}}

\newpage

Figure 10:\\   
\vspace{1in}
\centerline{\epsfxsize=6in\epsfbox{Figures/fig10.epsi}}
\newpage

\end{document}